\documentclass[sigplan,screen]{acmart}

\AtBeginDocument{%
  }

\copyrightyear{2026}
\acmYear{2026}
\setcopyright{cc}
\setcctype{by}
\acmConference[ASPLOS '26]{Proceedings of the 31st ACM International Conference on Architectural Support for Programming Languages and Operating Systems, Volume 2}{March 22--26, 2026}{Pittsburgh, PA, USA}
\acmBooktitle{Proceedings of the 31st ACM International Conference on Architectural Support for Programming Languages and Operating Systems, Volume 2 (ASPLOS '26), March 22--26, 2026, Pittsburgh, PA, USA}
\acmPrice{}
\acmDOI{10.1145/3779212.3790210}
\acmISBN{979-8-4007-2359-9/2026/03}

\usepackage{quantikz}
\usepackage[]{hyperref}
\usepackage{algpseudocode}
\usepackage{yquant}
\usepackage[nameinlink,capitalise]{cleveref}
\usepackage{verbatim}
\usepackage{amsmath}
\usepackage{amsthm}
\usepackage{amsfonts}
\usepackage{braket}
\usepackage{bbm}
\usepackage{bm}
\usepackage{xcolor}
\usepackage{xspace}
\usepackage{tcolorbox}
\usepackage{graphicx}
\usepackage{subcaption}
\usepackage{enumitem}
\usepackage{multirow}
\usepackage{microtype}[final]
\usepackage{bbm}

\usepackage{fancyhdr}
\usepackage[normalem]{ulem}
\usepackage{listings}
\usepackage[noend,linesnumbered,ruled,vlined]{algorithm2e}
\usepackage{physics}

\usepackage{amssymb}
\newcommand{\ignore}[1]{}
\usepackage{adjustbox}
\usepackage[normalem]{ulem}
\usepackage[flushleft]{threeparttable}
\usepackage{dblfloatfix}

\usepackage{subcaption}
\usepackage{color,soul}
\usepackage{balance}

\setlist{nolistsep}
\useyquantlanguage{groups}

\definecolor{aliceblue}{rgb}{0.97, 0.96, 1.0}

\newenvironment{mybox}[1][aliceblue]{  %
    \begin{tcolorbox}[   %
        left=0pt,
        right=0pt,
        top=0pt,
        bottom=0pt,
        colback=#1,
        colframe=#1,
        width=0.99\dimexpr\columnwidth\relax,
        boxsep=2pt,
        arc=0pt,outer arc=0pt,
    ]
}{
    \end{tcolorbox}
}

\newcommand{\rev}[1]{{{\color{black}#1}}}

\newcommand{\ours}{\textsc{trasyn}\xspace}
\newcommand{\gridsynth}{\textsc{gridsynth}\xspace}
\newcommand{\synthetiq}{\textsc{Synthetiq}\xspace}
\newcommand{\bqskit}{\textsc{BQSKit}\xspace}
\newcommand{\pyzx}{\textsc{PyZX}\xspace}

\begin{document}
\pagenumbering{gobble}

\title{\rev{Reducing T Gates with Unitary Synthesis}}

\author{Tianyi Hao}
\orcid{0000-0003-4074-4971}
\affiliation{%
  \institution{University of Wisconsin-Madison}
  \city{Madison}
  \state{WI}
  \country{USA}}
\email{tianyi.hao@wisc.edu}

\author{Amanda Xu}
\orcid{0009-0008-2279-5816}
\affiliation{%
  \institution{University of Wisconsin-Madison}
  \city{Madison}
  \state{WI}
  \country{USA}}
\email{axu44@wisc.edu}

\author{Swamit Tannu}
\orcid{0000-0003-4479-7413}
\affiliation{%
  \institution{University of Wisconsin-Madison}
  \city{Madison}
  \state{WI}
  \country{USA}
}
\email{swamit@cs.wisc.edu}

\begin{abstract}

Quantum error correction is essential for achieving practical quantum computing but has a significant computational overhead. Among fault-tolerant (FT) gate operations, non-Clifford gates, such as $T$, are particularly expensive due to their reliance on magic state distillation. These costly $T$ gates appear frequently in FT circuits as many quantum algorithms require arbitrary single-qubit rotations, such as $R_x$ and $R_z$ gates, which must be decomposed into a sequence of $T$ and Clifford gates. In many quantum circuits, $R_x$ and $R_z$ gates can be fused to form a single $U3$ unitary. However, existing synthesis methods, such as \gridsynth, rely on indirect decompositions, requiring separate $R_z$ decompositions that result in a threefold increase in $T$ count.

This work presents \rev{\em TensoR-based Arbitrary unitary SYNthesis (\ours)}, a novel FT synthesis algorithm that directly synthesizes arbitrary single-qubit unitaries, avoiding the overhead of separate $R_z$ decompositions. By leveraging tensor network-based search, our approach enables native $U3$ synthesis, reducing the $T$ count, Clifford gate count, and approximation error. Compared to \gridsynth-based circuit synthesis, for 187 representative benchmarks, our design reduces the \textit{T count} by up to \textbf{3.5}$\times$, and \textit{Clifford gates} by \textbf{7}$\times$, resulting in up to \textbf{4}$\times$ improvement in overall circuit infidelity.
\end{abstract}




\maketitle

\section{Introduction}\label{sec: intro}

Quantum error-correcting (QEC) codes enhance the reliability of quantum computing using redundant information, by encoding a single logical qubit into multiple physical qubits. Although this enables fault-tolerant quantum computing (FTQC), it significantly slows down computation. For instance, computations on these logical qubits are executed through logical gates; however, the runtime and resource cost of these gates vary significantly. Particularly, the \(T\) gate is $100\times$ slower compared to physical gates for most QEC codes due to a procedure known as magic state distillation~\cite{bravyi2005universal}. Magic state distillation uses multiple logical qubits to generate a highly purified (nearly error-free) magic state, essential for implementing a single \(T\) gate. Since each logical qubit generally comprises hundreds of physical qubits, the overhead grows quickly with the number of \(T\) gates.

This overhead is further exacerbated by the limited gate set available within QEC codes~\cite{knill2005} and the need for gate decomposition. For example, the \(R_z(\theta)\) gate, common in quantum circuits, cannot be directly executed in FTQC. It must instead be decomposed into a series of simpler gates supported by the FTQC. Typically a combinations of Clifford gates \{\(H\), \(S\), \(X\), \(Y\), \(Z\)\} and \(T\) gates are used to emulate \(R_z(\theta)\) gate~\cite{gridsynth,kitaev1997quantum}.

\[
R_z(\theta) = \begin{bmatrix}
e^{-i\theta/2} & 0 \\
0 & e^{i\theta/2}
\end{bmatrix} \xrightarrow{\text{Decompose}} H T Z T H X \dots H
\]

This required decomposition significantly increases the number of $T$ gates and thereby demand for magic states, which are necessary to perform fault-tolerant $T$ gates. These $T$ gates primarily dictate the execution time of the quantum algorithms ~\cite{reiher2017elucidating,beverland2020lower,beverland2022assessing}. In this paper, we introduce a \textit{circuit synthesis method that reduces the number of T and Clifford gates} to reduce resource overhead to help us realize fault-tolerant quantum algorithms sooner.

\rev{Most existing FT synthesis methods are based on number theory and rely on the diagonal structure of $R_z$~\cite{selinger2012efficient, kliuchnikov2013asymptotically, kliuchnikov2015practical, gridsynth, bocharov2013efficient, kliuchnikov2015approximateFramework, kliuchnikov2023shorter}. 
In particular, \gridsynth~\cite{gridsynth} is considered state-of-the-art and is broadly adopted as a subroutine by 
FT studies and compilers~\cite{nam2020approximate, resch2021benchmarking, choi2023fault, blunt2024compilation, stein2024architectures, watkins2024high, wang2024optimizing}.
}
\gridsynth provides optimal or near-optimal decompositions for $R_z$ gates at any given approximation error threshold. However, $R_z$ gates represent only a subset of all possible single-qubit unitaries. A general single-qubit unitary must be decomposed into three $R_z$ rotations:
\begin{equation}
\begin{footnotesize}
U3(\theta,\phi,\lambda) = R_z(\phi+\frac{5}{2}\pi) H R_z(\theta) H R_z(\lambda-\frac{\pi}{2}), \label{eq: u3-decomp}
\end{footnotesize}
\end{equation}

which requires three $R_z$ gates to implement a single arbitrary unitary ($U3$) gate. This results in \textbf{a threefold increase in $T$ count} compared to directly synthesizing $U3$.\footnote{The three \( R_z \) decompositions may exceed the error budget (\(\epsilon\)), requiring it to be lowered to \(\frac{\epsilon}{3}\), which increases \( \#T \) by more than \( 3\times \).
}\footnote{\rev{\cite{kliuchnikov2023shorter} improved the $3\times$ overhead to $\frac{7}{3}\times$ by associating the three syntheses.}}

\begin{figure}[t]
    \centering
\includegraphics[width=\columnwidth]{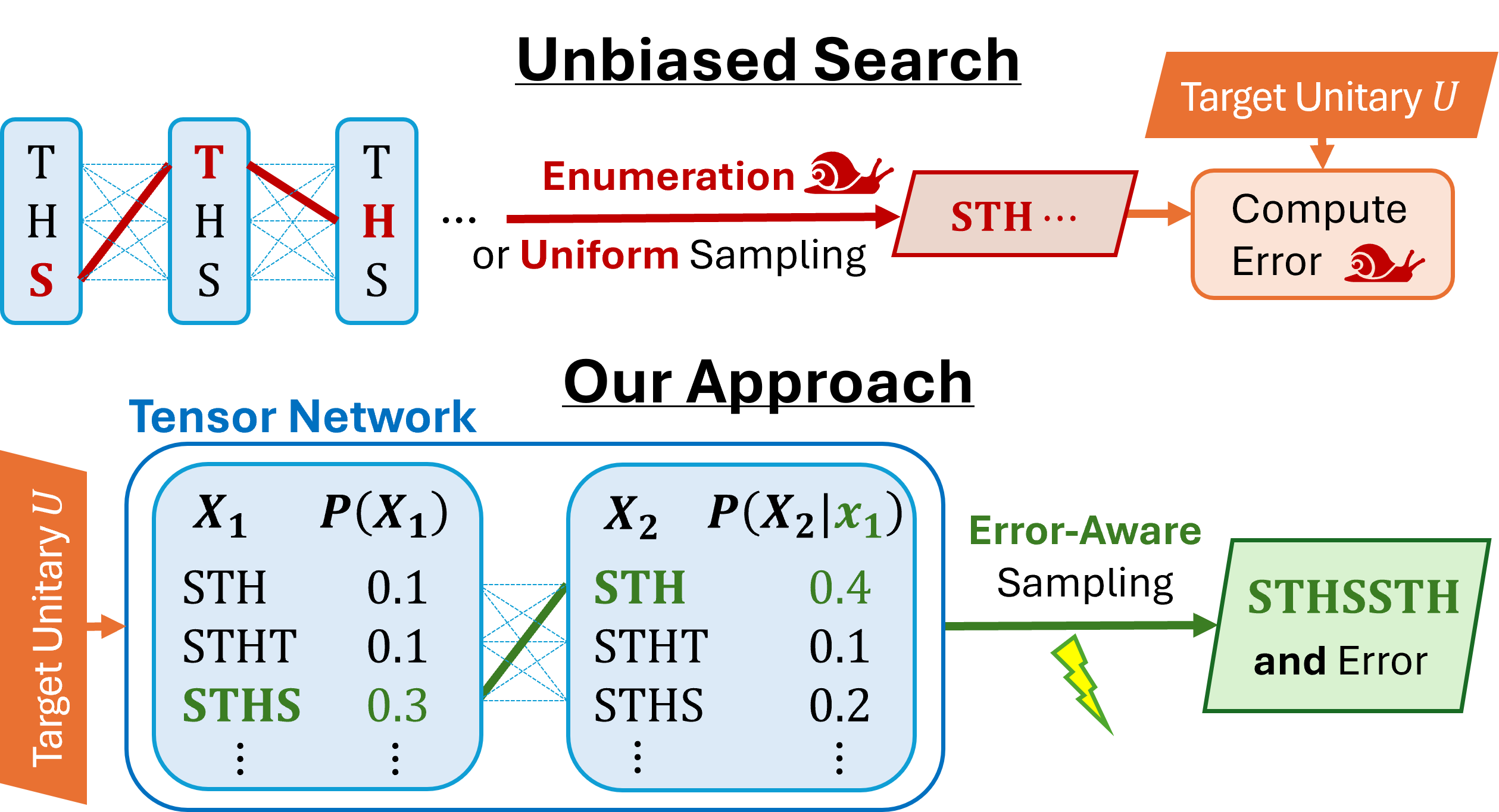}
\caption{(a) Brute-force enumeration or sampling of gate sequences is slow and unscalable due to the vast search space. (b) Our method \ours leverages precomputed sequences as building blocks to extend the scale beyond brute force. By incorporating the target unitary and interpreting the error (distance to the target) as probabilities, \ours enables error-aware sampling, delivering efficiency and accuracy.}
\label{fig:intuition}
\vspace{-0.1in}
\end{figure}

In this paper, we demonstrate for the first time that many quantum circuits provide opportunities to merge multiple $R_z$ and $R_x$ gates into single $U3$ gates, leading to significant reductions in both $T$ count and total gate count. Despite these advantages, the approach of transpiling into Clifford+$U3$ and directly synthesizing $U3$ gates has not been widely explored. The primary challenge is that general unitary synthesis is considerably more difficult than $R_z$ synthesis, as it lacks the simplifications offered by the diagonal structure of $R_z$ gates.

In arbitrary unitary synthesis, the goal is to find the best sequence of gates, selected from a discrete set of gates such as Clifford+\( T \) that closely matches the target operation \( U \). 
Unfortunately, existing unitary synthesis methods often produce low-quality solutions~\cite{dawson2005solovay,kitaev1997quantum} or are impractical due to their high computational demands in terms of time and memory~\cite{fowler2011constructing, paradis2024synthetiq}. Our work addresses these limitations by enabling efficient unitary synthesis that significantly reduces the number of $T$ gates and the overall gate count, leading to faster execution times and improved scalability for FTQC.

\rev{

\paragraph{Our approach} Figure~\ref{fig:intuition} shows our key insight: we encode the exponentially large solution space as a compact \emph{tensor network}. Our approach {\em TensoR-based Arbitrary unitary SYNthesis (\ours)} constructs a matrix product state (MPS) from short, precomputed gate sequences. This network implicitly represents longer sequences and supports efficient computation of trace distances via structured tensor contractions. By attaching the target unitary to the network, we can construct a probability distribution over the search space weighted by distance—enabling \emph{error-aware sampling} that preferentially selects high-quality sequences.

This tensor-guided approach offers two key benefits over brute-force or uniform sampling: (1) we avoid repeatedly computing the distance—each sample comes with its error “for free,” as the tensor network representation enables \emph{concurrent evaluation of trace distance across exponentially many sequences}, and (2) we gain fine-grained control over the synthesis process, such as enforcing $T$-count budgets or reusing modular building blocks. This allows us to scale synthesis well beyond the limits of exhaustive methods while producing lower-$T$ and higher-fidelity circuits.

\vspace{0.1in}
{\small
\centering
    \begin{tabular}{l|c|c|c}
    \footnotesize
    \textbf{Method} & \textbf{Strategy} & \textbf{Synthesis Error} & \textbf{Scale} \\
    \hline
    Brute Force & Exhaustive & $10^{-2}$ & $\lesssim$15 $T$ gates \\
    \ours{} & Tensor-guided & $10^{-4}$ & 30+ $T$ gates \\
    \end{tabular}
}
\vspace{0.1in}
}

Compared to prior methods such as \gridsynth or \synthetiq, \ours achieves significantly lower $T$ counts and better scalability, making it practical for early FT.

\begin{figure}[t]
    \centering
    \includegraphics[width=\linewidth]{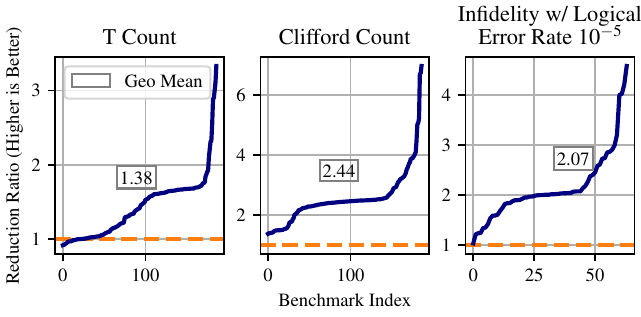}
    \caption{Reduction ratio ($\frac{\gridsynth}{\ours}$) for $T$ count, Clifford gates, and infidelity for baseline to our design. Values of more than one indicate our design performs better.}
    \label{fig:intro}
    \vspace{-0.15in}
\end{figure}

We evaluate \ours for 1000 randomly selected single-qubit unitary operators, \ours reduces the $T$ count by \rev{$3.7\times$} compared to \gridsynth with similar approximation errors. Furthermore, we use representative circuit benchmarks to compare 
\gridsynth-based $R_z$ workflow against the $U3$ workflow enabled by \ours.
Figure~\ref{fig:intro} shows reductions in the \( T \) gate count, Clifford gate count, and improvement in fidelity across 187 representative benchmarks compared to \gridsynth, including quantum chemistry and material simulations, and quantum optimization algorithms from established quantum benchmark suites~\cite{nation2024benchmarking,mqtbench,qasmbench,hamlib,Supermarq}. 

Our evaluation demonstrates that \ours achieves a geometric mean reduction in $T$ count of \textbf{1.38}$\times$ and up to \textbf{3.5}$\times$ (as highlighted in Figure~\ref{fig:intro}). Additionally, our design significantly decreases the number of $S$ gates in synthesized sequences, leading to a maximum reduction of \textbf{7}$\times$  in Clifford gates for these benchmarks compared to \gridsynth. 

Moreover, we study the tradeoff between synthesis and logical errors. Synthesis errors arise from approximating \( U3 \) with a limited \( T \) gate budget, while logical errors occur when QEC fails to correct faults. Reducing synthesis error requires more \( T \) gates, increasing the overall gate count and the likelihood of logical errors. Our analysis shows that allowing a slightly higher synthesis error—by using fewer \( T \) gates—can improve overall fidelity by minimizing logical error accumulation. When accounting for realistic logical error rates (on the order of  $10^{-5}$) anticipated in early fault-tolerant (EFT) systems~\cite{EFT1,EFTQAOA,EFT2, EFT3}, our design demonstrates up to \textbf{4}$\times$ improvement in overall circuit infidelity.

With tensor networks, we enable scalable circuit synthesis, a critical step toward practical FTQC. Our contributions are:

\begin{itemize}

\item We propose \ours, a \textit{generic} and \textit{efficient} single-qubit unitary synthesis algorithm that uses tensor networks to guide the search. Our implementation is amenable to GPU acceleration and outperforms state-of-the-art synthesis algorithms.

\item We demonstrate that using a $U3$-based intermediate representation significantly reduces rotations and the corresponding $T$ count compared to the widely accepted \( R_z \)-based alternative.

\item \rev{We identify the optimal synthesis error threshold given a logical error rate} and show that \ours achieves a synthesis error sufficiently small for EFT logical error rates such as \(10^{-7}\).

\end{itemize}

\section{Background and Motivation} %

\subsection{Restricted Fault-Tolerant Gate Sets}

No quantum error-correcting code can implement a universal gate set transversally, as transversal gates act independently on each physical qubit~\cite{eastin2009restrictions}. Consequently, at least one non-Clifford gate, typically the \(T\) gate, must be implemented using resource-intensive techniques such as magic state distillation~\cite{bravyi2005universal,bravyi2012magic,haah2017magic}.  

Magic state distillation refines multiple noisy copies of \(\ket{T^*}\) into a smaller set of high-fidelity magic states, enabling fault-tolerant \(T\) gate implementation. 
This process relies on logical ancilla qubits for error detection, measurement, and state projection to achieve the required fidelity. 
However, this process significantly increases the physical qubit overhead and execution time. Since magic state generation is the primary bottleneck in FT quantum computation, the resource cost of executing an FT algorithm is often estimated by the \(T\) gate count in the decomposed circuit~\cite{gidney2021factor,reiher2017elucidating}. Parallel magic state factories, which are a dedicated collection of logical qubits executing the distillation protocol—can produce large numbers of magic states, mitigating this bottleneck through a space-time tradeoff. However, near-term FT quantum computers will likely be constrained by the limited availability of physical qubits, resulting in slow computation. Therefore, reducing the \(T\) count through efficient compilation techniques is critical for improving performance.

\subsection{Standard FTQC Compilation Workflows}
\begin{figure}[t]
    \centering
    \includegraphics[width=0.63\linewidth]{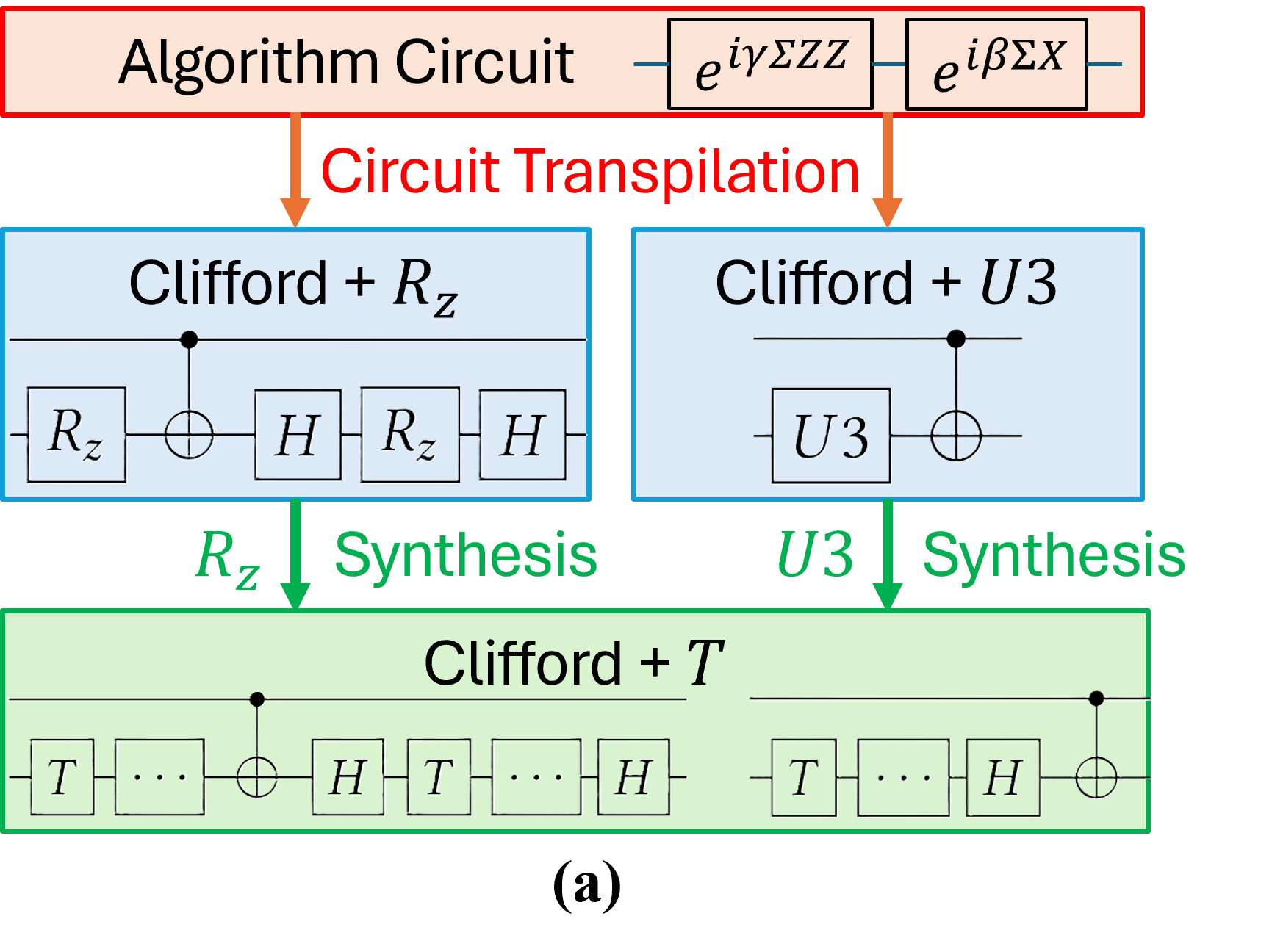}
    \includegraphics[width=0.36\linewidth]{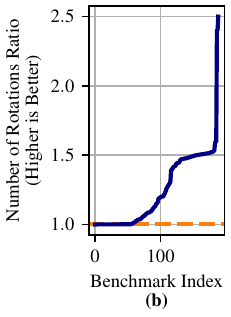}
    \caption{(a) Compilation workflows for FTQC. The $U3$ IR enables the merging of multiple rotations and can lead to fewer $T$ gates after synthesis. (b) Ratio of \(R_z\) to \(U3\) gate counts after transpiling 187 benchmark circuits into $CNOT$+$H$+\(R_z\) and $CNOT$+\(U3\), indicating the effectiveness of $U3$ as IR. }
    \label{fig:intro_work}
\end{figure}

To execute quantum circuits on an FTQC architecture, complex unitary operations must be decomposed into basic gates compatible with the error correction protocol, typically the Clifford+$T$ gate set. Figure~\ref{fig:intro_work}(a) illustrates a quantum circuit synthesis workflow where the given unitary operator used by the algorithm is first transformed into a circuit using an intermediate representation (IR) with either Clifford+$R_z$ or Clifford+$U3$ gate set. This intermediate circuit is then converted into Clifford+$T$ gates through single-qubit synthesis protocols. The choice of IR dictates the synthesis algorithm used for the final decomposition into Clifford+T gates.
We study the question: \begin{center}\textbf{Which IR is better for FT compilation: $U3$ or $R_z$?}\end{center}

To address this question, we analyzed 187 representative circuits by decomposing them into two target gate sets, $U3$ and $R_z$, using the Qiskit transpiler at four optimization levels (0, 1, 2, 3). For each gate set, we counted the number of {rotations}\footnote{Unless otherwise noted, we concern ``nontrivial'' rotations - rotations requiring more than one $T$ gate - given the fact that $R_z$ angles of integer multiples of $\frac{\pi}{4}$ can be synthesized exactly with zero or one $T$ gate while other angles require a substantial amount of $T$ gates to synthesize to a reasonable accuracy~\cite{kliuchnikov2012fast}.} when decomposed. We pick the optimization level with minimum rotations.  We then calculated the ratio of $R_z$ to $U3$ rotations, as shown in Figure~\ref{fig:intro_work}(b). Our evaluations show that many benchmarks show opportunities for merging multiple gates into a single $U3$ to reduce the number of rotations, which can lead to a lower total $T$ gate count compared to the $R_z$-based workflow. However, current state-of-the-art single-qubit decomposition methods only support $R_z$ synthesis, forcing users to adopt a suboptimal gate set despite $U3$'s advantages.

\subsection{Arbitrary Single-Qubit Unitary Synthesis}\label{sec: arbitrary-unitary-synthesis}
The main reason $U3$ is not adopted is because existing unitary synthesis methods produce low-quality solutions, are slow, or do not scale\rev{~\cite{dawson2005solovay, fowler2011constructing, paradis2024synthetiq, alam2023quantum}}.
The \textit{Solovay-Kitaev algorithm}~\cite{dawson2005solovay} was the first method for approximating arbitrary single-qubit unitaries using fault-tolerant gate sets. For a target error \( \epsilon \), it produces gate sequences of length \( O(\log^c(1/\epsilon)) \), where \( c > 3 \). This is far from the \rev{information-theoretic lower bound of \( K + 3\log_2(1/\epsilon) \)~\cite{selinger2012efficient} for ancilla-free synthesis, where $K$ is a constant.} Giving the algorithm more resources, such as time, also does not improve solution quality. 
Fowler introduced an enumeration-based method\rev{~\cite{fowler2011constructing}} that achieves a lower \( T \) gate count than Solovay-Kitaev but suffers from exponential runtime, limiting its practical utility. In the category of techniques that will produce higher-quality solutions when given more time, Synthetiq~\cite{paradis2024synthetiq} is the state-of-the-art synthesis algorithm for discrete gate sets. However, their simulated-annealing approach struggles to scale to practical error thresholds, as we show in \cref{sec:eval}, while our approach \emph{can} scale.

\subsection{Problem Statement} 
The goal of this paper is to enable practical single-qubit synthesis of arbitrary unitary operators using discrete gate sets.
We select the Clifford+$T$ gate set as an example, comprising \(\{H, S, T, X, Y, Z\}\) matrices. Our goal is to construct a sequence of these gates such that their matrix product \( V\) approximates the target \(U\) to a specified accuracy. Two key metrics define synthesis quality. The \emph{T count} (\(\#T\)) measures the number of \(T\) gates (equivalently, non-Clifford Gates in other gate sets). As \(T\) gates are resource-intensive in error-corrected systems and dominate overhead, minimizing \(T\) count is our primary focus. The second metric is the closeness between the synthesized operator $V$ and the target $U$. 
We measure the closeness based on the Hilbert-Schmidt inner product $|\Tr(U^\dag V)|/N$,
    where $N$ is the dimension of $U$ ($N=2$ in our single-qubit case). 
    In the rest of the paper, we refer to it as the \emph{trace value} 
    and do not explicitly write $N$. 
    We extend it to measure the \emph{unitary distance}:
    \begin{equation}
    \small
        D(U, V)=\sqrt{1 - |\Tr(U^\dag V)|^2/N^2},\label{eq: gpid2}
    \end{equation}
    which is numerically very close to the operator norm $\left\lVert U - V \right\lVert$ for small values that appear as errors in this paper.~\footnote{The operator norm is widely used as the error metric by number-theoretic synthesis methods, such as our baseline \gridsynth. 
    }

    Formally, we define two forms of the FT gate synthesis problem. Given Clifford gate set $S_c$, non-Clifford gate set $S_{nc}$, unitary matrix $U\in \mathrm{U}(2)$, error threshold $\epsilon\in[0,1)$, and $T$ budget $\#T$, we aim to find a list of matrics $G=[g_i]$, where $g_i\in S_c\cup S_{nc}$, that will
    \begin{equation}
        \mathrm{minimize}\ D(U, \textstyle\prod_{g\in G} g)\ \mathrm{subject\ to}\ \sum_{g\in G}\mathbbm{1}_{S_{nc}}(g)\le\#T.\label{eq: problem1}
    \end{equation}
    Alternatively, the objective may be formulated as
    \begin{equation}
        \mathrm{minimize}\ \textstyle\sum_{g\in G}\mathbbm{1}_{S_{nc}}(g)\ \mathrm{subject\ to}\ D(U, \textstyle\prod_{g\in G} g)\le \epsilon.\label{eq: problem2}
    \end{equation}

\section{\ours: A Single-Qubit Unitary Synthesis Method for Discrete Gate Sets}

\subsection{Overview}\label{sec: method-overview}

   Given a discrete gate set (Clifford+$T$) and a specified $T$ gate budget ($\#T$), our goal is to synthesize a sequence of gates from the gate set, whose product closely approximates a target unitary $U$. We propose \ours, an efficient synthesis method that leverages tensor networks to systematically construct such gate sequences. 
   At a high level, \ours calculates the trace values $\Tr(U^\dag V)$ for every possible $V$ that can be constructed with up to $\#T$ gates and finds a gate sequence with a high trace value. 
   However, the number of possible $V$ scales exponentially in $\#T$.
    Explicitly enumerating every possible $V$ and calculating all the trace values is time-consuming at compile time and becomes infeasible beyond $\#T>15$. These limitations make the naive trace-based approach intractable and inadequate in the accuracy it can reach. We address the former by utilizing precomputation and lookup tables, avoiding unnecessary computation during compilation. We address the latter by concatenating multiple precomputation results and implicitly calculating the trace values. 

    \rev{Intuitively, we have a large set of matrices and are searching for the closest one to the target. The higher accuracy we require, the larger this set becomes. Because enumerating one monolithic set is intractable, we compose matrices from smaller sets by multiplying them to construct a solution close to the target. This approach naturally involves a large number of matrices and their products, which form tensors and tensor networks. Tensor networks can provide a compact representation of a large discrete space, in our case, the unitaries exactly representable by the gate set. They also offer established numerical methods for us to utilize and adapt, such as the sampling algorithm.
    }

\subsection{Tensor Networks}\label{sec: tensor-networks}
    \begin{figure}
        \centering
        \includegraphics[width=\linewidth]{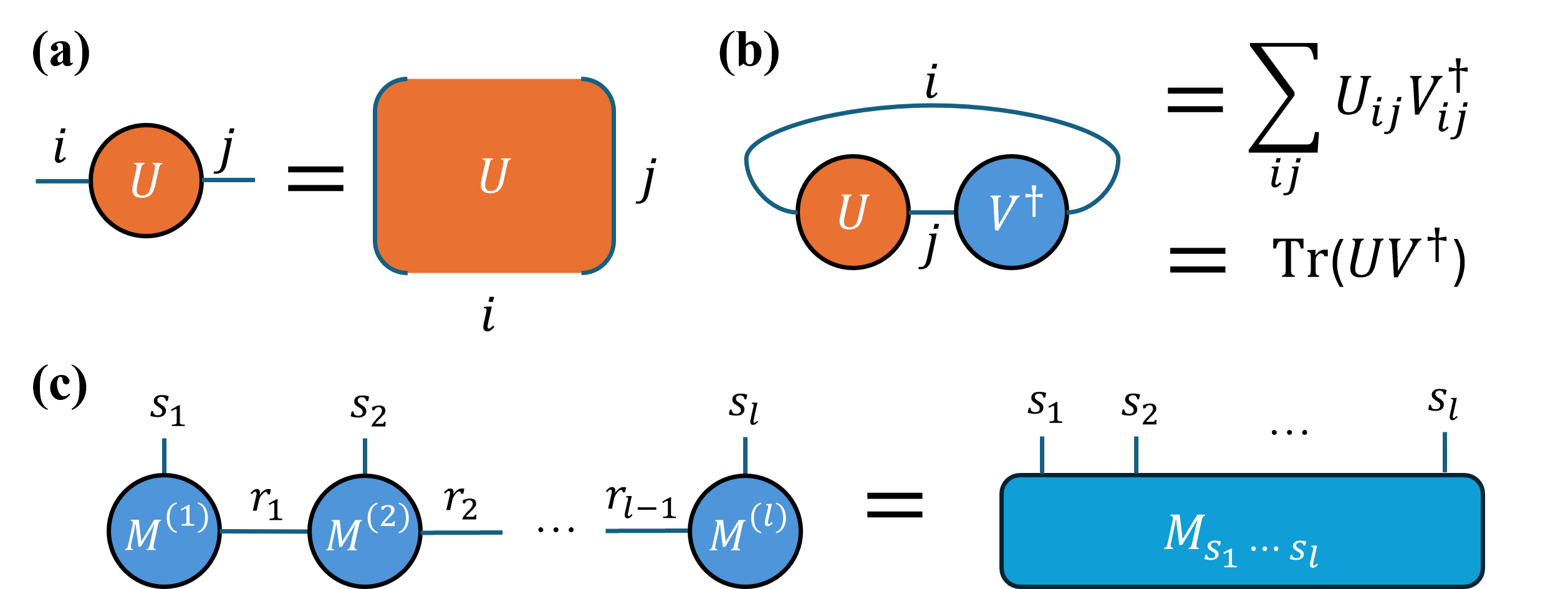}
        \caption{(a) A matrix drawn in tensor diagram. Each dimension becomes an edge of the tensor node. (b) The trace operation on two matrices is to sum over the two dimensions. (c) A 1D tensor network (MPS) and the equivalent exponentially large tensor.}
        \label{fig: tensor}
    \end{figure}
    
    Tensors are multi-dimensional arrays that generalize scalars, vectors, and matrices.
    Tensor contractions are the generalization of matrix multiplications, which are performed by summing over shared dimensions (also known as bonds or indices) between two tensors.
    Tensor diagrams are an intuitive way of visualizing tensors, where we use nodes to represent tensors and edges to represent dimensions, as shown in \Cref{fig: tensor}~(a). \Cref{fig: tensor}~(b) shows the tensor diagram of taking the trace $\Tr(U V^\dag)$, a crucial operation in our method.
    
    A tensor network is a generalized graph of tensors connected by shared dimensions, which is a powerful computational tool widely used in Mathematics, Physics, Chemistry, and Computer Science~\cite{orus2014practical, vidal2003efficient, ibrahim2022constructing, he2021high, pang2020efficient, hao2022quantum, tang2022scaleqc}. 
    In particular, \emph{matrix product states (MPS)}~\cite{schollwock2011density}, also known as tensor train~\cite{oseledets2011tensor}, are very successful in various applications~\cite{sidiropoulos2017tensor,chan2011density, hao2024variational}. 
    An $l$-site MPS consists of two 2-dimensional tensors $M^{(1)}_{s_1 r_1},M^{(l)}_{s_l r_{l-1}}$ and ($l-2$) 3-dimensional tensors $M^{(2)}_{s_2 r_1 r_2},\cdots,$\\$M^{(l-1)}_{s_{l-1} r_{l-2} r_{l-1}}$, such that the full contraction
    \begin{align}
        \sum_{r_1 r_2 \ldots r_{l-1}}M^{(1)}_{s_1 r_1}M^{(2)}_{s_2 r_1 r_2}\cdots M^{(l-1)}_{s_{l-1} r_{l-2} r_{l-1}} M^{(l)}_{s_l r_{l-1}} \label{eq: mps}
    \end{align}
    represents an $l$-dimensional tensor $M_{s_1,\cdots,s_l}$ (\Cref{fig: tensor}~(c)). These uncontracted dimensions are called \emph{physical indices}.
    
    In our context, \ours constructs an MPS that represents the exponentially large tensor of trace values of every possible $V$, where the physical indices are associated with gate sequences. Thus, we can use an algorithm that finds the indices corresponding to high trace values in the MPS to identify high-quality gate sequences without forming the exponentially large tensor. 
    \rev{Note that this differs from the usual MPS construction commonly seen in quantum physics, where physical indices represent quantum system dimensions. Our construction lets us utilize MPS to efficiently represent the exponentially many candidate unitaries and the corresponding gate sequences, even if we only work with single-qubit 2-by-2 matrices. This is in contrast to the usual use of MPS to efficiently represent the exponentially large Hilbert space of high-dimensional quantum systems.
    } 
    
    \begin{figure*}
        \centering
        \includegraphics[width=\linewidth]{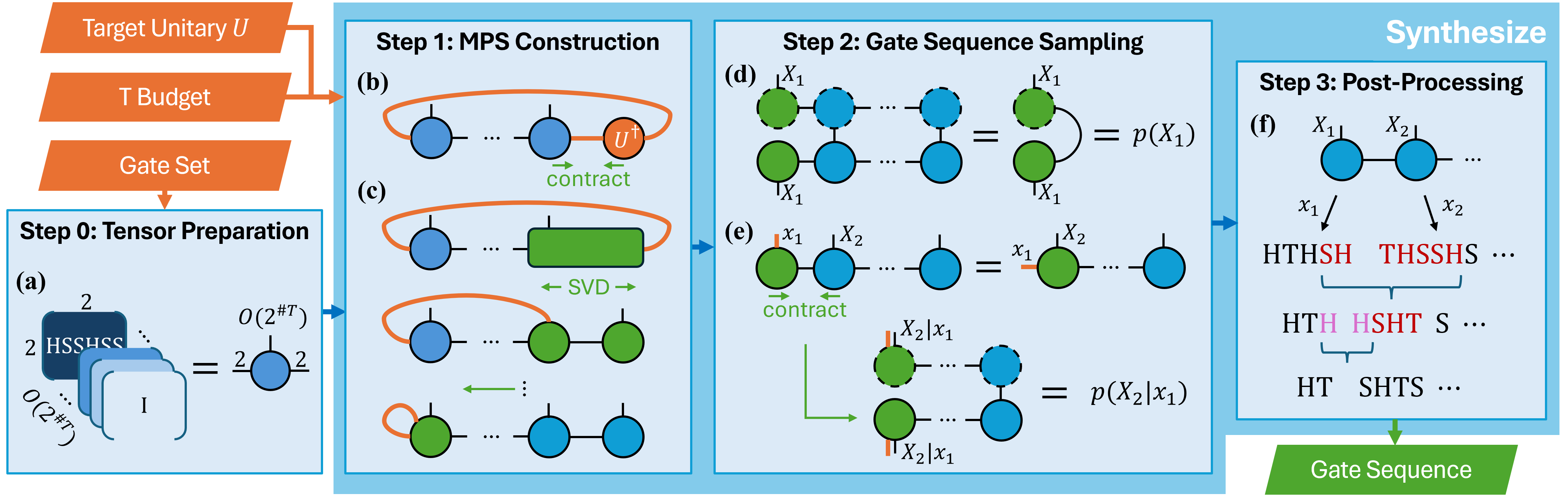}
        \caption{An overview of \ours's core algorithm. (a) Step 0 identifies unique matrices and the associated gate sequence. and stores them as tensors. Squares represent the matrices stacked together, which becomes a 3-dimensional tensor, where the extra dimension indexes the matrices and is our sampling target. (b) To go beyond brute-force enumeration, step 1 connects the tensors to construct an MPS and implicitly traces the target unitary along the matrix dimensions (orange). (c) Sequential contractions and decompositions (denoted in green) are performed to shift the matrix dimensions to the same tensor to perform the trace operation. The same process is repeated toward the left for each pair of tensors. (d) Step 2 samples gate sequences using the trace values as a joint probability distribution $p(X_1,\ldots,X_l)$, where each $X_i$ represents the sequence to choose for each tensor. To compute $p(X_1)$, we sum over $X_2,\ldots,X_l$, which is a contraction with the conjugate transposes, denoted as dashed nodes. (e) We sample a value $x_1$ from $p(X_1)$ and project the first tensor to $x_1$, contract with the second tensor, and repeat the same process to compute $p(X_2|x_1)$. (f) Step 3 further optimizes the sampled gate sequence. Although sequences are prepared to be optimal for each tensor, the concatenation of sequences from multiple tensors can contain suboptimal subsequences with shorter equivalents.}
        \label{fig: workflow}
    \end{figure*}

\subsection{Method}\label{sec: method}
    \ours consists of four steps (shown in \Cref{fig: workflow}): a preparation step that only needs to be done once per gate set and three steps performed at compile time. Step 0 prepares the tensors and gate sequences to be used in the MPS; step 1 constructs the MPS and implicitly performs the trace operation with the target unitary; step 2 samples the indices of high-quality solutions from the MPS based on the trace values; step 3 post-processes the sampled gate sequence to further reduce $T$ count and total gate count. We describe each step in detail in the following paragraphs.

\paragraph{Step 0: tensor preparation.}
    Step 0 is a one-time precomputation, where we enumerate all unique (up to a global phase) single-qubit matrices within a fixed $T$ count budget and the shortest gate sequences to produce them.
    In the following, we use a basis set of $\{T, S, H, X, Y, Z\}$, but the algorithm can be applied to other discrete gate sets.
    
     We start by identifying unique matrices given by Clifford gates and store the shortest (in $S$ and $H$ counts) sequences that produce them. Then, for each $T$ count below the budget, we enumerate new sequences with one more $T$ gate by taking the previously found sequences, adding one $T$ gate, and concatenating with the Clifford sequences. For each new sequence, we use the trace value to check if its matrix is already found. We store new matrices and the corresponding sequences. For duplicates, we compare the new sequence with the stored sequence and store the one with smaller $T$, $S$, and $H$ counts (order depends on gate cost assumptions). Once all unique matrices up to the fixed $T$ count have been identified, we stack them into a 3-dimensional tensor where the extra dimension indexes the different matrices. \Cref{fig: workflow}~(a) illustrates this construction, where the squares represent the unique matrices, each with a corresponding gate sequence.
     During duplication checking, we also generate a lookup table that records equivalent sequences, allowing us to later replace longer sequences with their shorter equivalents.

    We find that there are $24 \cdot (3 \cdot 2^{\#T} - 2)$ unique single-qubit matrices for up to $\#T$ $T$ gates. 
    Our finding agrees with the theoretically proven result of $192 \cdot (3 \cdot 2^{\#T} - 2)$
    ~\cite{matsumoto2008representation}, which counts the 8 possible global phases allowed by the Clifford+$T$ gate set. 
    The cost of enumerating all unique matrices thus scales exponentially with $T$ count and quickly becomes intractable after approximately 15 T gates.~\footnote{Our optimized implementation took days on an NVIDIA A100 GPU to enumerate unique matrices with 15 \(T\) gates, scaling as \(O(4^{\#T})\). This is a one-time cost as the FT gate set is fixed.} To enable the synthesis of more $T$ gates, we proceed to steps 1 and 2 with the tensors and sequences we have prepared from enumeration.

\paragraph{Step 1: MPS construction.}
    Step 1 extends our synthesis capabilities beyond 15 $T$ gates by linking the prepared tensors along their $2\times2$ matrix dimensions to form an MPS. 
    During this process,
    we select tensors with specific $T$ count ranges based on the overall $T$ budget. 
     By specifying the number of tensors in the MPS and each tensor's $T$ count range, we can control the $T$ count range for synthesizing the target unitary.

    We then attach the target unitary and perform the implicit trace operation. 
    The normal trace operation on two matrices is a summation over the two dimensions they share. 
    To trace the target unitary $U$ with our constructed MPS, we connect (but not sum over yet) its two dimensions with the two matrix dimensions at the beginning and the end of the MPS (\Cref{fig: workflow}~(b)). To eliminate the loop that would hinder efficient sampling, we perform sequential contractions and singular value decompositions (SVDs) to shift the target’s second dimension from the end to the beginning of the MPS (\Cref{fig: workflow}~(c)). This effectively contracts the matrix dimensions of the MPS with those of the target unitary. The resulting MPS implicitly captures all the trace values, so that when the MPS is fully contracted, its entries are the desired trace values, ready for the sampling process in step 2.
    
    For $l$ tensors, each with $m$ $T$ gates, the time complexity of step 1 is $O(2^m l)$, dominated by the $l$ contractions and SVDs on tensors of size $O(2^m)$.

\vspace{0.05in}

\paragraph{Step 2: sampling gate sequence.}
In step 2, we sample high-quality gate sequences based on the trace values implicitly captured by the MPS. We adopt the existing MPS sampling technique, which is traditionally used to efficiently sample the basis states of quantum systems~\cite{ferris2012perfect} and values of multivariate functions~\cite{optima_tt,chertkov2023tensor, dolgov2020approximation, chertkov2022optimization}. Adapting to our context, we interpret the trace values as a joint probability distribution over all indices. By applying the chain rule, we decompose this joint distribution into a product of conditional probabilities:
    \begin{equation}
    p(X_1, \ldots, X_l) = p(X_1) ~p(X_2|x_1) ~\cdots~ p(X_l|x_1, \ldots, x_{l-1}),
    \end{equation}
    where each $X_i$ represents one open physical index (i.e., one part of the gate sequence).
    The sequential SVDs in step 1 bring the MPS to the so-called canonical form, where only one tensor is non-isometric. The other tensors contract to the identity with their conjugate transpose, allowing the conditional probabilities to be computed locally, beginning with the first tensor.

    To compute $p(X_1)$, we sum over all other indices $X_2, \ldots, X_l$, which is equivalent to locally computing a partial inner product with the conjugate of the first tensor (\Cref{fig: workflow}~(d)). More memory-efficiently, we can calculate this distribution by taking the 2-norm of the first tensor along the index connecting to the second tensor. Once we have $p(X_1)$, we sample a value $X_1=x_1$ and then compute $p(X_2|x_1)$.

    Given the chosen $x_1$ value, we project the first tensor to that value by simply slicing out that index, such that the remaining problem becomes sampling from $p(X_2, \ldots, X_l | x_1)$. We then contract with the second tensor and repeat the same process to compute $p(X_2|x_1)$ (\Cref{fig: workflow}~(e)). This procedure continues through all tensors to produce one complete set of indices $(x_1, x_2, \ldots, x_l)$ sampled from the joint distribution. To be more efficient, we obtain multiple gate sequences in one pass by selecting multiple indices at each distribution sampling, where each index may correspond to different previous selections. We denote the number of samples as $k$.

    The complexity of step 2 is $O(2^m kl)$, resulting from a single pass of $l$ local tensor operations, where $O(2^m)$ comes from the size of the tensor.

    \paragraph{Step 3: post-processing} In step 3, we loop over the sampled gate sequence to identify subsequences that can be shortened. Although sequences are prepared to be optimal for each tensor, step 2 gives a concatenation of sequences from multiple tensors, which can lead to opportunities for further optimization, as shown in \Cref{fig: workflow}~(f). Thus, we substitute suboptimal subsequences with shorter equivalents using the lookup table we enumerated in step 0.

    Overall, our method takes $O(2^m kl)$ time and $O(2^m k)$ space to sample $k$ gate sequences with $T$ count up to $ml-2$.

\paragraph{Full Algorithm}
    The algorithm consists of steps 1 to 3 (we denote as \texttt{Synthesize()}) solves the synthesis problem in the form of \Cref{eq: problem1}. In particular, the algorithm prioritizes lowering the error within a $T$ budget and reports the best solution instead of solutions closer to the thresholds. We extend \ours's capability to solve the problem in the form of \Cref{eq: problem2} by wrapping subroutine \texttt{Synthesize()} in an outer loop, as specified in Algorithm \ref{alg: trasyn}. 
    We provide an option to supply an error threshold, where \ours reattempts the synthesis when the threshold is not satisfied. 
    We allow the $T$ budget to be specified as a list, where each tensor can have a different $T$ count range. We attempt the synthesis from a minimum number of tensors $l$ to the full $T$ budget list, ensuring we start from a low $T$ budget first and incrementally increase the total budget in searching for better solutions. If we set the budget increments to be 1, Algorithm~\ref{alg: trasyn} effectively solves \Cref{eq: problem2}.

\paragraph{Implementation}
    We provide an implementation of \ours in Python, with \texttt{NumPy} as the only dependency. We offer optional GPU acceleration based on \texttt{CuPy}. We have made significant efforts on engineering optimizations, ensuring \ours to be performant in practice. \ours is open-source, available at \hyperlink{https://github.com/haoty/trasyn}{https://github.com/haoty/trasyn}.

{\small
\setlength{\textfloatsep}{0pt}
\SetAlgoNoLine
\begin{algorithm}[t]
\caption{\ours}\label{alg: trasyn}
\SetKwInput{KwInput}{Input}                %
\SetKwInput{KwOutput}{Output}              %
\DontPrintSemicolon
  
    \KwInput{ $U$ (\textrm{target unitary}), $m$ (\textrm{list of $T$ budgets}), $l$ (\textrm{minimum number of tensors}), $k$ (\textrm{number of samples}) $\epsilon\gets\textsc{null}$ (\textrm{error threshold}), $r$ (\textrm{number of attempts})}\vspace{0.0 in}
    \KwOutput{ $s$ (\textrm{gate sequence}), $e$ (\textrm{synthesis error})}\vspace{0.05 in}

  \SetKwFunction{TRASYN}{TRASYN}
  \SetKwFunction{SYNTHESIZE}{SYNTHESIZE}
 
  \SetKwProg{Fn}{Function}{:}{\KwRet}
  \Fn{\TRASYN{$U, m, l, k, \epsilon, r$}}
  {
      \For{$i = l,\ldots, |m|$ \textbf{and} $j = 1,\ldots, r$}{
            Sequence $s$, Error $e = \mathrm{Synthesize}(U, m[:I], k)$\;
            \lIf{$e < e_{\text{best}}$}{$s_{\text{best}},e_{\text{best}} \gets s,e$}
            \lIf{$(\epsilon\neq\textsc{null}) \land (e < \epsilon)$}{\KwRet $s_{\text{best}},e_{\text{best}}$}
    }
    \KwRet $s_{\text{best}},e_{\text{best}}$\;    
  } %
\end{algorithm}}

\subsection{Application Specific Optimization with \ours}\label{sec: transpile}
We discuss example scenarios where $U3$ leads to fewer rotations. Trivially, all adjacent single-qubit gates can be merged as a single $U3$. As an example, hardware-efficient ansatzes for the variational quantum eigensolver often have adjacent axial rotations. In some FT algorithms, two $R_z$ gates separated by an $H$ gate can also be merged as a $U3$. A slightly advanced scenario is when two-qubit gates, such as $CNOT$s, separate single-qubit gates that they can commute with. For example, $R_z$ and $R_x$ commute with $CNOT$ on the control and target qubit, respectively. As an example of utilizing these commutations, we can merge all but one $R_x$ per layer in the quantum approximate optimization algorithm (QAOA) circuits (with quadratic Hamiltonians) into adjacent $R_z$ gates. For 3-regular graphs, these merges create a 40\% reduction in the number of rotations, which can be consistently obtained, as we will show in \Cref{sec: circuit-benchmark}.

\begin{figure}[t]
    \centering
    \includegraphics[width=0.8\linewidth]{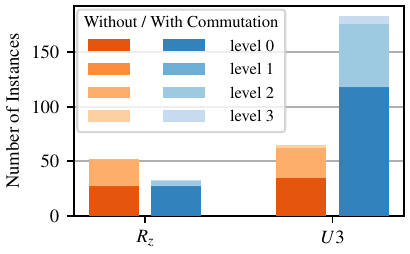}
    \caption{Number of instances a transpilation setting leads to the least number of rotations. The $U3$ IR benefits from the commutation pass and, in most cases,  leads to circuits with fewer rotations to be synthesized.}
    \label{fig: transpile}
\end{figure}

Beyond manually constructible scenarios, we have to rely on circuit transpilers. We employ the widely-used \texttt{Qiskit} transpiler and compare $R_z$ and $U3$ IR with four optimization levels. In addition, we add the option to run the gate commutation pass, 
which is not included in the default four levels of transpilation workflows.
\rev{This pass commutes $R_z$  with the control of a $CNOT$ and $R_x$  with the target of a $CNOT$.}
\Cref{fig: transpile} shows the number of times a particular transpilation setting results in the least number of rotations across all transpilation settings. 
We see that $U3$ enables the flexibility of the commutation pass and leads to fewer rotations to be synthesized for most circuits. Across all circuits, $U3$ brings an up to $2.5\times$ reduction in the number of rotations.

\section{Evaluation}
\label{sec:eval}

We designed our evaluation of \ours to answer the following research questions:
\begin{description}
    \item[RQ1] How does \ours compare to state-of-the-art tools when synthesizing arbitrary single-qubit unitaries?
    \item[RQ2] Is the error threshold \ours can scale to meaningful?
    \item[RQ3] How does \ours perform against state-of-the-art gate synthesis tools for application circuits?
    \item[RQ4] What is the impact of \ours on application fidelity?
    \item[RQ5] Can \ours's advantage be reclaimed by the state-of-the-art post-synthesis circuit optimization tool?
\end{description}

\paragraph{Metrics} We consider three resource metrics when comparing against other methods: (1) $T$ count, (2) $T$ depth, defined as the $T$ count on the critical path of a circuit, (3) Clifford count, where we exclude Pauli gates since they are free in error-correcting codes~\cite{chamberland2018fault,knill2005}. These metrics suffice to give an idea of the impact on circuit execution time without making assumptions about the exact latencies of each operation, which can vary depending on the architecture and error-correcting code. We also consider two error metrics: (1) unitary distance (\Cref{eq: gpid2}), and (2) state infidelity $1-\lvert\langle\psi_{\text{synth}}|\psi_{\text{true}}\rangle\rvert^2$. 
We compute the unitary distance for circuits under 12 qubits and estimate the state infidelity for circuits under 27 qubits in the ideal case and 12 qubits in the noisy case.

When comparing against baseline tools, we plot the \emph{metric ratio} between \ours and each baseline. Any values above 1 indicate \ours outperforms the baseline, and values equal to 1 or below 1 indicate \ours matches or underperforms the baseline, respectively. For example, the $T$ count ratio is defined as $\frac{\text{Baseline T count}}{\text{\ours T count}}$. Unless otherwise noted, we take the $\log$ for the error metrics, following the inverse logarithmic relationship between $T$ count lower bound and the error threshold. To maintain the consistency that a higher reduction ratio is better, we also invert the ratio for $\log$ values: $\frac{\log(\text{\ours error})}{\log(\text{Baseline error})}$. Since both the $T$ count and the error cannot be exactly fixed, we primarily aim to make the errors from different tools at around the same level (error ratios close to 1) and compare the $T$ counts.

\rev{\paragraph{Baselines} We compare \ours to state-of-the-art tools related to fault-tolerant circuit compilation and optimization, including gate synthesis tools \gridsynth~\cite{gridsynth} and \synthetiq~\cite{paradis2024synthetiq}, circuit partitioning and resynthesis tool \bqskit~\cite{younis2021bqskit}, and $T$ count optimization tool \pyzx~\cite{kissinger2019pyzx}. Our primary baseline, \gridsynth,  is broadly adopted as a subroutine to synthesize $R_z$ gates into Clifford+$T$ by a wide range of studies, including FT demonstrations, resource estimations, and architectural improvements~\cite{nam2020approximate, resch2021benchmarking, choi2023fault, blunt2024compilation, stein2024architectures, watkins2024high, wang2024optimizing}.
\synthetiq can synthesize arbitrary unitaries, including multi-qubit ones. Both tools are capable of approximate synthesis.
At the circuit level, \bqskit partitions a circuit into small subcircuits and resynthesizes them by numerical instantiation. \bqskit outputs circuits with $R_z$ gates, requiring further syntheses using \gridsynth.
\pyzx optimizes the $T$ count for circuits already in Clifford+$T$.
}

\paragraph{General experimental setup} \ours uses an NVIDIA A100-80GB GPU \rev{ with \texttt{CUPY\_ACCELERATORS=cub}}, and other tools use 24 CPU cores on an AMD EPYC 7763 machine and 80GB of RAM. 
We enable \gridsynth's flag to find solutions up to the global phase.
\rev{We modify \synthetiq's error metric to be consistent with \Cref{eq: gpid2}.}

\subsection{RQ1: Single-Qubit Unitary Synthesis}\label{sec: random-unitary}
\paragraph{Benchmarks} We sample 1000 single-qubit unitary matrices uniformly at random from the Haar measure, which ensures an unbiased set of arbitrary single-qubit unitaries. 

\paragraph{Experimental setup} We allocate 10 minutes per unitary for \ours, \gridsynth, and \synthetiq. We explore several synthesis error thresholds relevant for EFT: 0.1, 0.01, and 0.001. \gridsynth only supports $R_z$ synthesis, so it requires a preprocessing pass to decompose the unitary into three $R_z$ rotations, as shown in \Cref{eq: u3-decomp}. We scale the error thresholds by $1/3$ for \gridsynth to account for the combined error from three syntheses.

\paragraph{Instantiation of \ours} We configure \ours to take 40,000 samples, allow 10 $T$ gates in each tensor, and have 1, 2, and 3 tensors in the MPS for the three target error thresholds, respectively.

\paragraph{Results}
    \begin{figure}[t]
        \centering
        \includegraphics[width=\linewidth]{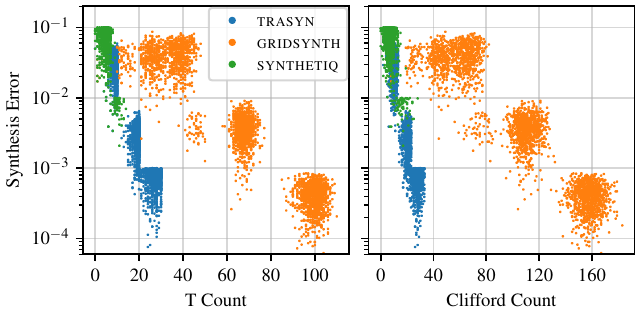}
        \caption{Synthesis error (unitary distance) and $T$ count of synthesizing 1000 random unitaries with \ours, \gridsynth, and \synthetiq at three different scales. \ours finds solutions using about $\frac{1}{3}$ $T$ gates and $\frac{1}{6}$ Cliffords compared to \gridsynth and outscales \synthetiq substantially. 
        }
        \label{fig: random-unitary}
    \end{figure}

    {\small
    \begin{table}
        \centering
        \caption{$T$ count and Clifford count reductions of \ours compared to \gridsynth for error threshold 0.001.}
        \begin{tabular}{c|ccccc}
        \toprule
        Reduction & Min & Mean & Geo Mean & Median & Max\\
        \midrule
        T Count & 2.31x & 3.76x & 3.74x & 3.68x & 6.12x\\
        \midrule
        Clifford Count & 3.39x & 5.77x & 5.73x & 5.66x & 9.41x\\
        \bottomrule
        \end{tabular}
        \label{tab: random-unitary}
    \end{table}
    }

    We plot the unitary distance between the synthesized matrix and the target matrix versus the $T$ count and the Clifford count as scatter plots 
    as all metrics have fluctuations in the results and cannot be fixed to a single value. In~\Cref{fig: random-unitary}, the three clusters of \gridsynth's results correspond to three error threshold settings, which exhibit the expected logarithmic relationship between the synthesis error and the $T$ count. \ours shows the same trend for results with $T$ budgets of 10, 20, and 30.

    Most notably, for the same level of error, \ours finds solutions using about $\frac{1}{3}$ $T$ gates compared to \gridsynth; and at about the same $T$ count, our synthesis errors are two orders of magnitude better. This comes from the fact that \gridsynth needs to synthesize for three $R_z$ rotations and concatenate the results, while our method can directly synthesize for the original unitary. We achieve a greater reduction in the non-Pauli Clifford count. \Cref{tab: random-unitary} summarizes the reductions we achieve at the 0.001 error threshold. 

   Compared to \synthetiq, \ours can reach orders of magnitude lower errors. \synthetiq cannot guarantee finding a solution within the time limit, even for the highest error threshold we ran. \synthetiq fails for 1, 931, and  1000 instances out of the total 1000 instances for error thresholds 0.1, 0.01, and 0.001, respectively. In fact, for the scale that \synthetiq can reliably finish, \ours can guarantee to find the optimal solution since only one tensor is needed, which effectively serves as a lookup table.
   Recall that \ours prioritizes lowering the error within a $T$ budget. For error thresholds 0.1 and 0.01, the $T$ budgets of 10 and 20 are more than enough to reach the error threshold, and \ours reports the best solutions found instead of solutions closer to the thresholds.

    Given that \ours is not an analytical algorithm, it cannot handle an arbitrarily small error threshold like \gridsynth does. The scale we show in \Cref{fig: random-unitary} is \ours's ``comfort zone'' with an A100 GPU, but not the limit. Given more time or memory, \ours can further scale to lower error levels.
    However, we will show in~\Cref{sec: logical-error} that an error threshold around 0.001 is optimal for logical error rates $10^{-6}$ to $10^{-7}$, under the most conservative assumption of the logical error.

\begin{figure}[t]
        \centering
        \includegraphics[width=\linewidth]{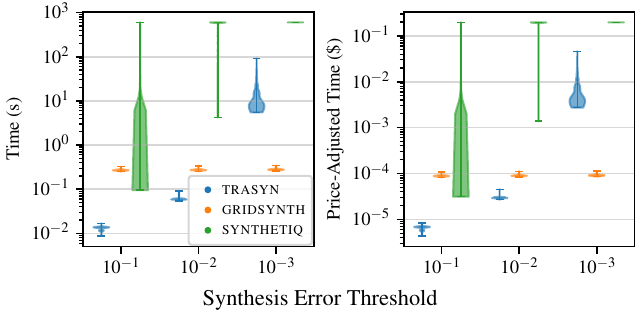}
        \caption{Time (s) and hardware-price-adjusted time (\$) for each method synthesizing 1000 random unitaries. \ours outperforms \gridsynth at 0.1 and 0.01 error thresholds and remains competitive at error threshold 0.001. \ours is significantly reliable than \synthetiq, which failed to find a solution by the 10-minute time limit for 1, 931, and 1000 instances out of the 1000 instances for each error threshold. }
        \label{fig: random-unitary-time}
    \end{figure}

   \Cref{fig: random-unitary-time} shows the synthesis time taken by each method. With GPU acceleration, \ours is even one to two orders of magnitude faster than the analytical method \gridsynth for error thresholds 0.1 and 0.01, enabling extremely fast run-time synthesis. At the threshold of 0.001, \ours still finishes most instances within seconds. Compared to \synthetiq, \ours's performance is much more stable and reliable. For most instances, \synthetiq hits the 10-minute time limit we set and cannot find a solution. To account for the hardware difference, we source the on-demand market pricing of the 24-CPU instance (\$1.18/hr~\cite{aws_pricing}) and one A100-80GB GPU (\$1.79/GPU/hr~\cite{lambdalabs_gpu_cloud}) and plot the price-adjusted time, which draws the same conclusions.

\begin{mybox}
    \paragraph{RQ1 summary} \textbf{For single-qubit unitary synthesis, \ours reduces $T$ count by \rev{$3.7\times$} compared to $R_z$ synthesis with \gridsynth and enables significantly faster and scalable $U3$ synthesis compared to \synthetiq.}
\end{mybox}

\subsection{RQ2: Logical and Synthesis Error Tradeoff}\label{sec: logical-error}

\begin{figure}[t]
    \centering
    \includegraphics[width=\linewidth]{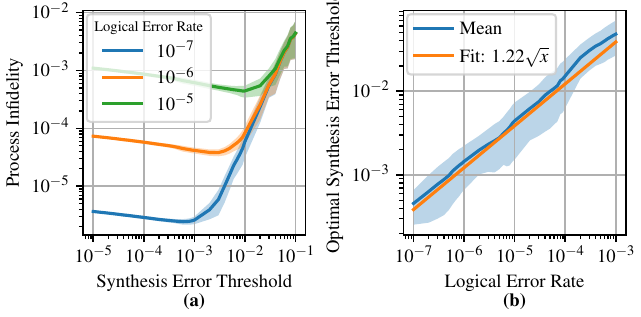}
    \caption{(a) Given a fixed logical error rate, there exists an optimal synthesis error threshold that minimizes the overall fidelity. (b) Fitting the optimal thresholds for varying logical error rates indicates a square root relationship. Shaded regions denote $\pm 1$ standard deviation.}
    \label{fig: tradeoff}
\end{figure}

\rev{Reducing the synthesis error leads to more $T$ gates in the decomposed gate sequence}, thereby requiring more QEC cycles to finish the execution, which can result in higher logical errors for a fixed code distance. By allowing a higher synthesis error, we can reduce the number of $T$ gates, thereby lowering the logical errors. We investigate this tradeoff between the synthesis error and the logical error, and its impact on the overall operational (process) fidelity.

\paragraph{Experimental setup} We decompose 1000 random $R_z$ gates with \gridsynth\footnote{We use \gridsynth because its solutions have theoretical guarantees on the closeness to the $T$ count lower bound.} under varying synthesis error thresholds from $10^{-1}$ to $10^{-5}$. We simulate each decomposed gate sequence under logical errors, which are modeled as depolarizing errors with error rates from $10^{-3}$ to $10^{-7}$. We only apply errors to $T$ gates, assuming Clifford gates are error-free, which represents a highly conservative model for logical errors \rev{and the worst-case scenario for the synthesis error}. We compute the process fidelity between the original $R_z$ and the product of decomposed gates under both synthesis and logical errors. This captures the realistic operational fidelity for measuring noisy quantum channels.

\begin{figure*}[t]
    \centering
    \includegraphics[width=\linewidth]{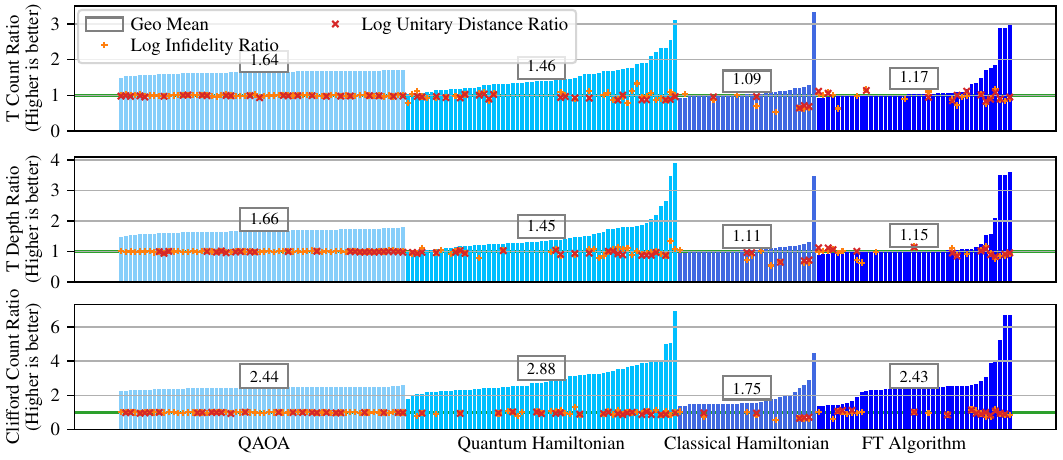}
    \caption{$T$ count, $T$ depth, and Clifford count reductions \ours achieves over \gridsynth across all benchmarks for about the same level of errors. Circuits with more diverse types of rotations, such as quantum Hamiltonians (with $R_x$, $R_y$, and $R_z$ instead of just $R_z$ in classical Hamiltonians), benefit more from $U3$ compilation and synthesis. With a circuit construction that enables maximum rotation merging, QAOA can consistently reach a high resource reduction.}
    \label{fig: gridsynth-comparison}
\end{figure*}

\paragraph{Results} \Cref{fig: tradeoff}~(a) shows the process infidelity as a function of synthesis error threshold for logical error rates $10^{-5}$, $10^{-6}$, and $10^{-7}$. As expected, we observe that given a fixed logical error rate, there is an optimal synthesis error threshold that minimizes the overall error. Setting the threshold higher or lower increases the process infidelity. We plot the optimal thresholds as a function of the logical error rate in \Cref{fig: tradeoff}~(b). 
The fitted function shows that the optimal synthesis threshold scales with the square root of the logical error rate. A threshold of 0.001 is sufficient for logical error rates between \(10^{-6}\) and \(10^{-7}\). For context, Google's latest error-corrected quantum computers~\cite{ai2024quantum} demonstrate a logical error rate of \(10^{-3}\), \rev{where the $T$ gate error rate can be several times higher due to magic state injection,} indicating that current hardware platforms are still far from making synthesis error a significant concern.
\rev{Furthermore, we note that an ideal FT synthesis method should be aware of the effects of logical errors and maximize the overall process fidelity when producing solutions. \ours's tensor approach is naturally suitable for such extensions.}

\begin{mybox}
    \paragraph{RQ2 summary}  \textbf{\ours is best suited for early FT platforms, with logical error rates of $10^{-7}$ and higher, where operational fidelity is constrained by the logical error rate rather than synthesis error.}
\end{mybox}

\subsection{RQ3: Circuit Benchmarks}\label{sec: circuit-benchmark}

Next, we explore how our improvements for synthesizing arbitrary single-qubit unitaries directly generalize to better Clifford+$T$ decompositions for application circuits. 

\paragraph{Benchmarks} We consider a diverse suite of 187 circuits, including standard FTQC algorithms, quantum chemistry and material simulations, and quantum optimization algorithms. Except for QAOA, we source the circuits from the Benchpress~\cite{nation2024benchmarking} and MQTBench~\cite{mqtbench} benchmarking repositories, which contain circuits from QASMBench~\cite{qasmbench} and Hamiltonians from Hamlib~\cite{hamlib}. We exclude circuits that are trivial to synthesize. For Hamiltonian simulation circuits, we employ state-of-the-art Pauli evolution gate compiler \textsc{Rustiq}~\cite{rustiq} to decompose multi-qubit Pauli terms into $CNOT$s and single-qubit gates. For QAOA, we generate 3-regular graph MaxCut circuits with a gate ordering that enables the maximum merge of rotations. We show the same reduction can be consistently achieved across QAOA depths and numbers of qubits. We vary QAOA depth from 1 to 5 and the number of qubits from 4 to 26. We generate 10 random instances with random angles for each QAOA circuit and take the average. 
\Cref{tab: benchmarks} summarizes the scope of the benchmarks in the number of qubits and the number of rotations. 
Before synthesizing individual rotational gates, we pass the circuits through Qiskit's transpiler with the setting leading to the fewest rotations from the 16 settings described in~\Cref{sec: transpile}.

{\small
\begin{table}
    \centering
    \caption{Datasets used in our full circuit benchmarks.}
    \begin{tabular}{c|ccc|ccc}
    \toprule
    \multirow{2}{*}{Dataset} & \multicolumn{3}{c}{\# Qubits} & \multicolumn{3}{c}{\# Rotations} \\
    \cmidrule(lr){2-4}\cmidrule(lr){5-7}
     & Min & Mean  & Max & Min & Mean  & Max \\
    \midrule
    Benchpress & 2 & 68.2 & 395  & 1 & 185.0 & 1531 \\
    \midrule
    Hamlib & 2 & 66.7 & 592  & 5 & 754.1 & 3875 \\
    \midrule
    QAOA & 4 & 15.0 & 26  & 6 & 73.4 & 209 \\
    \bottomrule
    \end{tabular}
    \label{tab: benchmarks}
\end{table}
}
    
\paragraph{Experimental setup}
We compare \ours on $CNOT$+$U3$ circuits against \gridsynth on Clifford+$R_z$ circuits. 
\rev{Each single-qubit $U3$ or $R_z$ is synthesized independently. The overall cost grows linearly with the number of single-qubit rotations in the circuit. These synthesizer invocations can be embarrassingly parallelized when compiling a large circuit with many rotations. The overall error can be bounded additively from individual rotation synthesis errors~\cite{patel_quest_2022}.}
To ensure the circuit-level errors of \ours and \gridsynth are close across benchmarks, we scale an original error threshold of 0.007 by the ratio of rotations between $U3$ and $R_z$ circuits for \gridsynth. The resulting effective error threshold will be lower than 0.007 to account for the accumulation of errors from having more rotations.

\rev{Additionally, we compare \ours against the \bqskit+ \gridsynth workflow, where \bqskit synthesizes the given circuit to Clifford+$R_z$ with the default configuration and \gridsynth synthesizes the $R_z$ gates with the same setup as described above.}

\paragraph{Instantiation of \ours}
We configure \ours to take 40000 samples and search from up to 20 $T$ gates.

\begin{figure}[t]
    \centering
    \includegraphics[width=\linewidth]{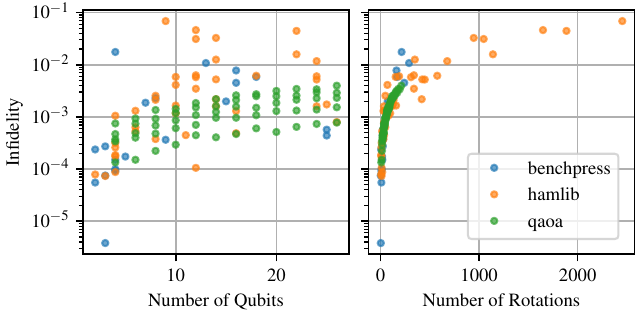}
    \caption{\rev{The actual circuit infidelity values synthesized by \ours in the circuit benchmarks, ordered by the number of qubits and rotations in the circuit, respectively.}}
    \label{fig: infidelity-values}
\end{figure}

\paragraph{Results}
\Cref{fig: gridsynth-comparison} presents the $T$ count, $T$ depth, and Clifford count reduction ratio \ours achieves over \gridsynth across all benchmarks for about the same level of \rev{overall circuit fidelities}.
\rev{In general, \ours achieves an average reduction in $T$ count of \textbf{1.39}$\times$ and up to \textbf{3.5}$\times$ across 187 benchmarks. These numbers deviate from the 3$\times$ reduction in single $U3$ experiments due to different amounts of rotation merges allowed in individual circuits.}
We separate benchmarks into different categories and observe that circuits with more diverse types of rotations benefit more from $U3$ compilation and synthesis. For example, Hamiltonians from classical problems only contain Pauli $Z$ terms, which transpiles to $R_z$ gates. Without other types of rotations in the circuit, $U3$ IR only achieves more rotation merges than $R_z$ IR in cases where two $R_z$ gates are separated by non-diagonal Cliffords. In contrast, Hamiltonians from quantum problems contain terms that are products of Pauli $X$, $Y$, and $Z$, which transpile to $R_x$, $R_y$, and $R_z$ gates, providing more rotation merging opportunities. FT algorithms also follow this pattern, where $R_z$-only algorithms typically only receive a small amount of $T$ count and $T$ depth reductions. Nonetheless, all benchmarks observe substantial Clifford reductions since \rev{ \gridsynth does not optimize for Clifford resources}.

As discussed in \Cref{sec: transpile}, QAOA serves as an example where careful construction of the circuits can lead to consistent rotation merges and thus, $T$ count reductions. Across the circuit sizes and QAOA depths we ran, \ours attains a $1.6\times$ $T$ count improvement, resulting from commuting the $R_x$ gates through $CNOT$s and merging with $R_z$ gates.

\rev{Since \Cref{fig: gridsynth-comparison} only shows the ratios, as a reference, \Cref{fig: infidelity-values} displays the actual values of the circuit infidelity resulting from the synthesis process.}

\begin{figure}[t]
    \centering
    \includegraphics[width=\linewidth]{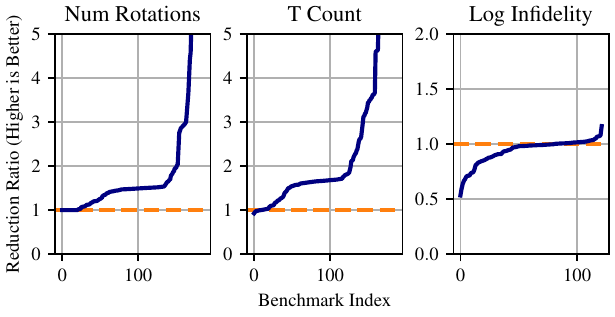}
    \caption{\rev{Reduction ratios of the number of rotations and $T$ count \ours achieves over \bqskit+ \gridsynth at around the same level of circuit infidelity.}}
    \label{fig: bqskit}
\end{figure}

\rev{Additionally, \Cref{fig: bqskit} compares \ours against the \bqskit+ \gridsynth workflow. We observe that \bqskit's numerical instantiation approach only increases the rotations, which in turn lead to more $T$ gates.}

\begin{mybox}
    \paragraph{RQ3 summary} \textbf{At the same error level, \ours outperforms state-of-the-art Clifford+$R_z$ transpilation and synthesis workflow across diverse application circuits in $T$ count, $T$ depth, and Clifford count.}
\end{mybox}

\subsection{RQ4: Impact on Application Fidelity}\label{sec: application-impact}
Previous section's extensive results demonstrate a broad scope of gate reductions enabled by \ours.
We now quantify the effect of gate reductions in the presence of logical errors in addition to the synthesis errors.

\paragraph{Experiment setup} We model the logical error as depolarizing errors of rates $10^{-4}$, $10^{-5}$, and $10^{-6}$ applied to all non-Pauli gates. These error rates reflect realistic near-term scenarios where limited physical qubits constrain QEC to smaller code distances. 
\rev{We use the empirical relationship in \Cref{fig: gridsynth-comparison} to derive the target synthesis error thresholds from the logical error rates, which are 0.0122, 0.00386, and 0.00122.}
We compute the fidelity for synthesized circuits under 12 qubits with density matrix simulation.

\begin{figure}[t]
    \centering
    \includegraphics[width=0.85\linewidth]{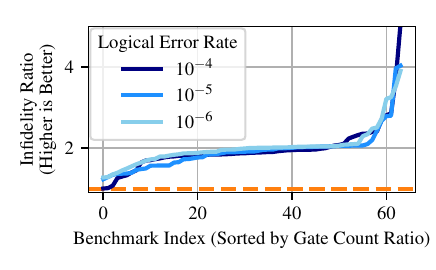}
    \caption{\rev{Infidelity ratio between circuits synthesized by \ours and \gridsynth under logical errors. \ours's gate count advantage leads to a higher application fidelity consistent across various logical error rates.}}
    \label{fig: gridsynth-logical-error-infidelity}
\end{figure}

\paragraph{Results} \Cref{fig: gridsynth-logical-error-infidelity} shows the infidelity ratio (without taking the $\log$) of the synthesized circuits under logical error rates $10^{-4}$, $10^{-5}$, and $10^{-6}$.
We order benchmarked circuits by total gate count ratio and observe that the reduction of synthesized gates leads to higher fidelity when logical errors are present. 
\rev{The advantages of \ours are consistent across different logical error rates, showing the benefits of $U3$ synthesis do not diminish with increasing logical errors.}

\begin{mybox}
    \paragraph{RQ4 summary} \textbf{Gate reductions achieved by \ours lead to higher circuit fidelities under logical errors, \rev{and the relative fidelity improvement stays persistent at different error rates.}}
\end{mybox}

\subsection{RQ5: Optimizing Synthesized Circuits}
Next, we investigate whether the improvements we observed in RQ2 remain after running a state-of-the-art $T$ count optimizer, \textsc{PyZX}~\cite{kissinger2019pyzx}, on the synthesized circuits. 

\paragraph{Experimental setup} We run \textsc{PyZX} on the circuits synthesized by \ours and \gridsynth from RQ2 that contain fewer than 50,000 gates. This limit is imposed to ensure \textsc{PyZX} can provide a solution in a reasonable amount of time while retaining a representative subset of the benchmark suite.  

\begin{figure}[t]
    \centering
    \includegraphics[width=\linewidth]{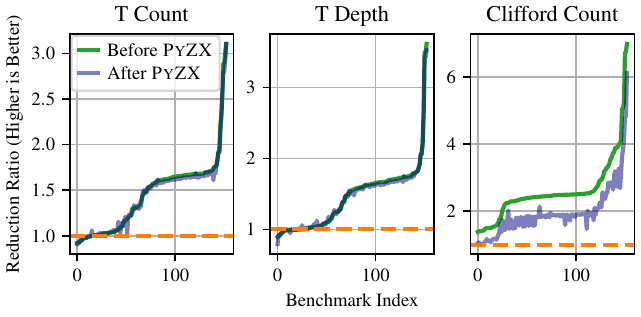}
    \caption{$T$ count, $T$ depth, and Clifford count ratios between \ours and \gridsynth before and after optimizing by \textsc{PyZX}. \textsc{PyZX} optimization is unable to level the $T$-advantages.}
    \label{fig: pyzx}
\end{figure}

\paragraph{Results}
\Cref{fig: pyzx} compares the $T$ count, $T$ depth, and Clifford count ratio of \ours and \gridsynth before and after optimization by \textsc{PyZX}. \textsc{PyZX} reduces the $T$ count and $T$ depth advantages of \ours by a small fraction for very few cases. For other few cases, these advantages are even slightly broadened. Although \textsc{PyZX} can reclaim some Clifford advantage, optimized \ours-synthesized circuits still have far fewer Clifford gates than optimized \gridsynth-synthesized circuits. Overall, \textsc{PyZX} is unable to reclaim the $T$-advantage of \ours, highlighting \ours's utility.

\begin{mybox}
    \paragraph{RQ5 summary}  \textbf{Synthesis is the primary contributor to \( T \) count, with post-synthesis optimizations like \textsc{PyZX} offering only marginal reductions.}

\end{mybox}

\section{Related Work}\label{sec: related-work}

\paragraph{Single-qubit synthesis}
Existing number-theoretic methods\rev{~\cite{selinger2012efficient, kliuchnikov2013asymptotically, kliuchnikov2015practical, gridsynth, bocharov2013efficient, kliuchnikov2015approximateFramework, kliuchnikov2023shorter}} can only synthesize $R_z$ rotations and need to perform at most three syntheses for a general unitary, which leads to 3$\times$ the $T$ count. In comparison, our method directly synthesizes the unitary and \rev{is general to different FT gate sets}. Compared to exhaustive enumeration methods, such as~\cite{fowler2011constructing}, our method utilizes precomputation and, thus, is much faster at runtime. Compared to methods that also take advantage of precomputation to build a table of matrices, our method can use these methods as our step 0 and be more scalable in $T$ count.

\rev{\paragraph{Multi-qubit synthesis} Multi-qubit synthesis techniques for continuous gate sets~\cite{younis2021bqskit, rakyta_approaching_2022} are more suited for reducing two-qubit gates and can often increase the number of arbitrary rotations, as shown in \Cref{sec: circuit-benchmark}. More related to this work, multi-qubit discrete synthesis methods~\cite{amy2013meet,giles2013exact,paradis2024synthetiq,kang2023modular,gheorghiu_t-count_2022, bqskit} are limited in scalability due to the significantly larger solution space. For example, the size of the Clifford group for $n = 1$, 2, and 3 is 24, 11,520, and 92,897,280, respectively. With $T$ gates involved, the number of possible unitaries under a given $T$ count is only going to grow more severely with the number of qubits. Consequently, multi-qubit synthesis is much less scalable in the $T$ count and usually focuses on optimizing frequently used FT gadgets with a few $T$ gates.

In principle, our method can target multi-qubit unitaries by including the two-qubit $CNOT$ gate in the gate set. Since unitary synthesis is an extremely hard problem, we limit ourselves to the single-qubit case and prioritize scaling $T$ count and synthesis error, which correlate to the logical error rate we can support.
}

\paragraph{Synthesis with additional resources} Prior methods leverage additional resources like ancillas, probabilistic circuits, or state distillation~\rev{\cite{bocharov2015efficientRUS, kliuchnikov2014asymptotically, akibue2024probabilistic,jones2013distillation, jones2012faster, duclos2013distillation, hastings2016turning, kliuchnikov2023shorter, bocharov2015efficientFallback, campbell2017shorter, mooney2021cost}} to reduce the $T$ count beyond the theoretical lower bound of resource-free approaches. 
\rev{The tradeoffs of allowing additional resources or directly implementing arbitrary rotations are an open research question as the assumptions and state-of-the-art of QEC rapidly evolve. 
In early fault-tolerant demonstrations, there is good reason to believe experimental researchers will target the vanilla Clifford + $T$ model first because of recent advancements in magic state cultivation for $T$ states~\cite{gidney2024magic}, whereas the equivalent does not exist for arbitrary rotation gates. Moreover, the direct distillation of arbitrary rotation often uses repeat-until-success protocols, which require complex control flow and very high-fidelity single-qubit gates, which may not suit all qubit architectures.
In addition, the early fault-tolerant regime will be extremely qubit-starved. Methods that reduce $T$ gates by using more logical qubits are likely less suitable for early FTQC.

On the other hand, direct decomposition methods are often foundational and complementary to methods that utilize more resources.
\ours requires no extra resources and is deterministic. We expect \ours to reach significantly lower errors if incorporated with these methods. For example, using \ours as a blackbox algorithm, mixing unitaries~\cite{campbell2017shorter, hastings2016turning, akibue2024probabilistic} can reduce the error quadratically. }

\paragraph{Quantum-circuit optimization} Unitary synthesis is closely related to circuit optimization. It is a core subroutine in some optimizers~\cite{younis2021bqskit, xu2024optimizingquantumcircuitsfast, Qiskit}. Thus, improving unitary synthesis has a direct impact on both circuit decomposition and optimization. Many circuit optimizers~\cite{xu2022synthesizing, xu2022quartz, li2024quarl, hietala2021verified} are better suited for reducing two-qubit gate count or total gate count, while some~\cite{kissinger2019pyzx, amy2024linearnonlinearrelationalanalyses, vandaele2024lowertcountfasteralgorithms, Heyfron_2019} solely focus on reducing $T$ count. Our approach is complementary to these optimizers. A reasonable workflow, as demonstrated in our evaluation, is to (1) run a general optimizer to reduce rotations, (2) use \ours to synthesize the rotations, and (3) run a $T$ count optimizer on the decomposed circuit. 

\section{Conclusion}

We present a novel FT synthesis algorithm for arbitrary single-qubit unitaries, eliminating the overhead of separate \(R_z\) decompositions. By leveraging a tensor-network-based search, our approach enables native \(U3\) synthesis, which compared to \gridsynth-based circuit synthesis, reduces \(T\) count by up to \textbf{3.5}$\times$, Clifford gates by \textbf{7}$\times$, and improve overall circuit fidelity by \textbf{4}$\times$. 

Furthermore, we analyze the tradeoff between logical and synthesis errors, showing that reducing synthesis error at the cost of a higher \(T\) count can worsen overall fidelity. Given the resource constraints of EFT quantum computing, this motivates a synthesis approach that prioritizes \(T\) count efficiency over achieving infinitesimal synthesis error. We demonstrate that \emph{arbitrary unitary} synthesis meets this need and is crucial for early fault tolerance.

\section*{Acknowledgement}
We sincerely appreciate the insightful discussions and valuable feedback from Aws Albarghouthi, Abtin Molavi, Ayushi Dubal, Meng Wang, Satvik Maurya, Zichang He, Zhixin Song, Cheng Peng, Yuchen Pang, and Yiqing Zhou. This research was supported by NSF Award \# 2340267 and \# 2212232. TH acknowledges the support from the JPMorganChase Ph.D. Fellowship. The numerical evaluations were conducted using computing resources from the Center for High Throughput Computing at the University of Wisconsin-Madison.

\bibliographystyle{ACM-Reference-Format}
\balance
\bibliography{references}

\end{document}